# Synthesis and properties of $SmO_{0.5}F_{0.5}BiS_2$ and enhancement in $T_c$ in $La_{1-y}Sm_yO_{0.5}F_{0.5}BiS_2$

Gohil Singh Thakur[§], Ganesan Kalai Selvan[¶], Zeba Haque[§], Laxmi Chand Gupta[§†], Saroj Lochan Samal[‡], Sonachalam Arumugam[¶] and Ashok Kumar Ganguli[§]*.

[§]Solid State and Nano Research Laboratory, Department of Chemistry, Indian Institute of Technology New Delhi 110016, India.

[¶]Centre for High Pressure Research, School of Physics, Bharathidasan University, Tiruchirapalli 620024, India

[‡]Department of Chemistry, National Institute of Technology, Rourkela, Odisha, 769008, India

## Abstract

Crystal structure and properties of a new member of oxy-bismuth-sulfide family $SmO_{0.5}F_{0.5}BiS_2$ are reported here. The compounds $SmO_{1-x}F_xBiS_2$ ($x$ = 0.0 and 0.5) are isostructural with $LaOBiS_2$ and crystallize in the $CeOBiS_2$ type structure ($P4/nmm$). Sm substitution in $LaO_{0.5}F_{0.5}BiS_2$, ($La_{1-y}Sm_yO_{0.5}F_{0.5}BiS_2$), leads to a gradual decrease in $a$-lattice constant however the $c$-lattice constant does not show such a gradual trend. Enhancement in $T_c$ is achieved upon partially substituting La by smaller Sm ion. Maximum $T_c$ ~ 4.6 K was observed for composition with $y$ = 0.8. Disobeying this trend $T_c$ disappears unexpectedly in composition $SmO_{0.5}F_{0.5}BiS_2$ ($y$ = 1.0). Both the undoped and F-doped ($x$ = 0.0 and 0.5) compounds are paramagnetic exhibiting semiconducting behavior down to 2 K.

# Introduction

Research in inorganic materials throws up unprecedented structures with unusual properties. The persistence of scientists carrying out exploratory research is challenging amidst dwindling recognition to this genre of scientists. Compounds like $Gd_2Cl_3$,[1] $NaMo_4O_6$,[2] $YBa_2Cu_3O_7$,[3] $K_3C_{60}$,[4] $TlCaBa_2Cu_2O_8$,[5] $CaCu_3Ti_4O_{12}$,[6] are compositions which rarely can be designed, but were synthesized by the unparalleled intuition (read ingenuity) and labour of solid state scientists which occasionally rewarded them with footprints for unusual material properties, like high temperature superconductivity, colossal magnetoresistance, unusually high dielectric constant, negative index of refraction, etc. This article follows from our work related to a recent class of superconductors, $Bi_4O_4S_3$ and $LnOBiS_2$ which have some structural similarity to the important LnOFeAs class of superconductors.[7-10] For example both these systems crystallize in the same space group (*P4/nmm*), they have layered structure with a charge reservoir layer (LnO) and a superconducting layer ($BiS_2$, Fe-As), their parent phases are semiconducting or poorly conducting and, most importantly, they become superconducting upon electron-doping (doping $F^-$ for $O^{2-}$ or a tetravalent ion for Ln).[9-17] Superconductivity by fluorine - doping in $LaO_{1-x}F_xBiS_2$ was first reported by Mizuguchi *et al* ($T_c \sim 2.5$ K)[9] ( $T_c \sim 10$ K could be achieved by high pressure synthesis).[9] Superconductivity in $LnO_{1-x}F_xBiS_2$ with Ln = Ce, Pr, Nd and Yb also followed soon.[10-16] $T_c$ is enhanced by chemical pressure[18-20] or external applied pressure[22-27] as observed for LnOFeAs[28] for example in $LnO_{1-x}F_xBiS_2$, a slight enhancement in $T_c$ has been reported by substituting Se at the S sites or by inter-mixing rare earth ions.[18-21] Hole doping via alkali earth-metal-substitution at Ln–site did not yield superconductivity.[22, 29] This is in contrast to what is observed in LnOFeAs where hole doping also induces superconductivity.[30-32] Electronic structure of these chalcogenides is comparable to that of pnictides and is greatly influenced by doping.[33-34] Their bands near the Fermi level consist of states predominantly arising from Bi ($6p$) and S ($3p$) states and therefore the Bi–S layers are responsible for superconductivity in these materials, while the LnO layer serves as the charge reservoir layer.[33-34] In this article we report: (1) the single crystal synthesis, structure and properties of a new member $SmO_{1-x}F_xBiS_2$. It has been found to be non-superconducting down to 2 K. (2) we have systematically investigated the effect of doping smaller Sm ions at La sites in $La_{1-y}Sm_yO_{0.5}F_{0.5}BiS_2$ and observed an enhancement in $T_c$ with Sm substitution.

**Experimental**

Polycrystalline samples of $La_{1-y}Sm_yO_{0.5}F_{0.5}BiS_2$ ($y = 0.0 - 1.0$) were synthesized by solid state method. $Ln_2S_3$, preheated $Ln_2O_3$, $LnF_3$, $Bi_2S_3$, Bi and S were ground well and pelletized in the form of circular disks. These disks were then sealed in evacuated silica tubes and heated at 800°C for 24 hours. The product was again ground, pelletized and heated at 800°C for 24 h for better homogeneity. The samples were stable in air but preserved in an Ar filled glove box to avoid any unnecessary contamination.

Single crystals of $SmO_{1-x}F_xBiS_2$ were grown using KCl-CsCl flux method. Polycrystalline $SmO_{1-x}F_xBiS_2$ was used as charge. Approximately 0.5g of compound was mixed with 2.5 g of KCl-CsCl (1:1). The properly mixed powder was sealed in an evacuated silica tube. The tube was heated at 800°C for 24 h and cooled to 600°C at a rate of 2°C/hour. Then the furnace was shut off and allowed to cool naturally. The tube was opened in ambient atmosphere and the flux was dissolved in distilled water. The product was washed several times with distilled water and finally rinsed with acetone. Many plate like micron sized (100-500 μm) crystals were obtained as observed under microscope. All the chemical manipulations except for vacuum sealing and washing of crystals were performed in an Ar filled glove box ($H_2O$ and $O_2$< 0.1 ppm).

The phase purity of all the samples was checked by powder X-ray diffraction technique using Cu-K$\alpha$ radiation in a *Bruker* D8 advance diffractometer. Single-crystal diffraction data sets were collected at room temperature over a 2θ range of ~4° to ~60° with 0.5° scans in ω and 10 s per frame exposures with the aid of a Bruker SMART CCD diffractometer equipped with Mo Kα radiation ($\lambda$ = 0.71073 Å). The data showed a primitive tetragonal lattice and the intensity statistics indicated a centrosymmetric space group. The reflection intensities were integrated with the APEX II program in the SMART software package.[35] Empirical absorption corrections were employed using the SADABS program.[36] The space group *P*4/*nmm* (No. 129) of the structures was determined with the help of XPREP and SHELXTL 6.1.[37] The structure was solved by direct methods and subsequently refined on $|F^2|$ with combinations of least squares refinements and difference Fourier maps. The refinements of $SmO_{0.5}F_{0.5}BiS_2$ converged to $R_1$ = 0.0662, $R_W$ = 0.1203 for all data with goodness of fit 1.17 and maximum residuals of 3.9 and −2.9 e/Å$^3$ that were 0.8 and 1.1 Å from Bi sites, respectively. Rietveld refinement analysis was carried out on all the samples using *Topas* software package.[38]

Magnetization measurements in the temperature range of 2 – 300 K on single crystals were performed by using a *Quantum Design* Superconducting Quantum Interference Device (SQUID) and on the polycrystalline samples by a vibrating sample magnetometer using a Physical Property Measurement System (PPMS). The resistivity studies in the temperature range of 2 – 300 K on polycrystalline samples were carried out by conventional four probe method using a Quantum Design PPMS.

**Results and discussion**

**1. $SmO_{1-x}F_xBiS_2$ ($x = 0.0, 0.5$)**

The Rietveld refined powder X-ray diffraction pattern of polycrystalline $SmO_{0.5}F_{0.5}BiS_2$ is shown in figure 1. Most of the peaks could be easily indexed on the basis of a tetragonal system with $CeOBiS_2$ structure in $P4/nmm$ space group. Presence of small amount (< 10 %) of $Bi_2S_3$ secondary phase was also detected which commonly occurs in these types of samples. The refined lattice parameters were found to be $a = 4.0122(5)$ Å, $c = 13.5949(2)$ Å for $x = 0.0$ and $4.0180(1)$ Å and $13.5291(4)$ Å for $x = 0.5$. Crystallographic data obtained from single crystal analysis for $SmO_{0.5}F_{0.5}BiS_2$ is given in Table 1. The corresponding atomic positions and isothermal displacement parameters are listed in Table 2. The CIF file and anisotropic displacement parameters are provided in the Supporting Information. It may be noted that due to similar scattering factors of O and F, their occupancies were not refined and were fixed to nominal values during refinement. $SmO_{1-x}F_xBiS_2$ is isostructural to $LaOBiS_2$ consisting of tetrahedral $Sm_2(O,F)_2$ and fluorite type $BiS_2$ layers (inset fig 1). The average stoichiometry of the F-doped single crystal sample was confirmed by EDAX analysis, results of which are given in Supporting Information. The relative ratio of Sm:F:Bi:S was found to be 1:0.39:0.96:2.1. The value of in-plane or equatorial Bi–S bond length ($Bi–S_{eq}$) was calculated to be 2.8423(2) Å and the S–Bi–S bond angle was 176.69 (4)°. A comparison of these bond lengths and bond angles of $Sm(O,F)BiS_2$ with other F- doped $LnOBiS_2$ including $Bi_4O_4S_3$ is shown in figure 2. It is to be noted that the value of 'x' in each LnO$_{1-x}$F$_x$BiS$_2$ is the one which gives the maximum $T_c$. The values for Ln = La,[39] (x = 0.46) and Nd[14] (x = 0.3) were obtained from the single crystal data reported elsewhere and for SmO$_{0.5}$F$_{0.5}$BiS$_2$ data from present studies were used. For Ce,[10] (x = 0.5), Pr[11] (x = 0.5) and Bi$_4$O$_4$S$_3$[7] the values were obtained from the data reported on polycrystalline samples. Since for Yb sample no structural refinement data has been reported, bond lengths and bond angles could not be calculated. An expected gradual decrease in both in–plane Bi–S bond distance and S–Bi–S bond angle is seen with decreasing size of the rare earth ion from La to Nd. For Sm system the Bi–S bond length is anomalously higher than Nd and which is a matter of further investigation. *Chen, et al,*[29] studied the effect of local distortion in Bi–S plane on the superconducting properties of La$_{1-x}$M$_x$(O,F)BiS$_2$ (where M = Mg and Ca). They find that a distortion in Bi–S plane is necessary to enhance superconductivity in LnOBiS$_2$. This seems to be true as on decreasing the size of rare earth ion, thereby increasing the distortion in Bi–S plane, $T_c$ is found to increase (from $T_c$ = 2.5 K in La to 5 K in Nd) in LnO$_{1-x}$F$_x$BiS$_2$. Similar enhancement in $T_c$ is also observed when two different rare earth ions are partially substituted for example in Ce$_{1-x}$Nd$_x$(O,F)BiS$_2$ and Nd$_{1-x}$Sm$_x$(O,F)BiS$_2$. $T_c$ increases with increase in substitution of smaller ion upto a certain value.[20-21]

Magnetization measurements in an applied field of 30 Oe carried out on a collection of large number of tiny single crystals (size ~ 100μm) of composition x = 0.0 and 0.5 exhibit a paramagnetic behavior down to 2 K (Fig. 3). These measurements clearly indicate absence of superconductivity at T > 2 K. The small hump at around 50 K is seen due to freezing of oxygen which might be present in the sample holder or due to slightly poor vacuum in the sample chamber. For the purpose of clarity, results of the variable temperature susceptibility (and its inverse) of the polycrystalline sample with x = 0.5 are shown in the low temperature range (T <

100 K) in figure 4. Curie constant 'C' obtained from the Curie-Weiss fit to the $\chi^{-1}$ vs T plot is C = 0.47 emu*K*mol$^{-1}$ and $\theta_p$ is –3.6 K. The Sm paramagnetic moment (as there is no other moment carrying species in the material) calculated using this value of C is 0.61 $\mu_B$. The deviation from the magnetic moment expected for free Sm$^{3+}$ ion (0.84 $\mu_B$) is possibly due to the crystal field effects. This magnetic moment is consistent with +3 ionic state of samarium. Field dependence of the magnetization is linear down to 5 K as shown in inset of Fig. 3, suggesting no magnetic ordering down to 5 K.

Resistivity measured for $x$ = 0.5 on a sintered polycrystalline pellet show a semiconductor like behavior similar to the undoped compound ($x$ = 0) (fig. 5). The resistivity value though decreased upon fluorine doping. We have not seen superconductivity down to 2 K in SmO$_{0.5}$F$_{0.5}$BiS$_2$ (see discussion in the next section). It may be pointed out that other members of this series are known to be superconducting (La, Ce, Nd, Pr and Yb) with $T_c$ ~ 2.5 K – 5 K.[15]

**La$_{1-y}$Sm$_y$O$_{0.5}$F$_{0.5}$BiS$_2$**

Compositions corresponding to La$_{1-y}$Sm$_y$O$_{0.5}$F$_{0.5}$BiS$_2$ were synthesized to investigate the possibility of superconductivity by doping La in non-superconducting SmO$_{0.5}$F$_{0.5}$BiS$_2$ and also study the effect of chemical pressure on $T_c$. Powder X-ray diffraction patterns of La$_{1-y}$Sm$_y$O$_{0.5}$F$_{0.5}$BiS$_2$ are presented in figure 6. Majority of the peaks correspond to the parent LaOBiS$_2$ phase along with some peaks for impurity phases like Bi$_2$S$_3$, Bi$_2$O$_3$, BiF$_3$ and/or LaF$_3$. Rietveld refinement on all the compositions was performed to study the crystal structure. The occupancies of La and Sm were refined constraining their sum to be unity. The occupancies of O and F were fixed to the nominal value. Results of Rietveld refinement and important structural parameters are provided in the Supporting Information. The variation of *a* and *c* lattice parameter with Sm doping ($y$) is presented in figure 7. There is a gradual decrease in the *a*-lattice parameter as we infer from the shift of the (200) peak towards higher angle (see figure S5 in supporting information). This is expected as Sm has a smaller ionic radius than that of La. The (004) peak does not show a gradual trend as a function of Sm-content (see figure S5 in supporting information). The inferred *c*-lattice parameter also, thus, does not show a gradual (smooth) variation as a function of Sm–content. This type of structural anisotropy is common in layered structures like Ln(O,F)BiS$_2$.[20-21] In contrast to the observation of Kajitani *et al*,[21] we do not find a solubility limit of Sm ions at La sites and hence, La was completely replaced by Sm and the sample with $y = 1.0$ *i.e.* SmO$_{0.5}$F$_{0.5}$BiS$_2$ could be obtained.

Superconductivity is observed in all the samples La$_{1-y}$Sm$_y$O$_{0.5}$F$_{0.5}$BiS$_2$ ($y = 0.0, 0.2, 0.4$ and $0.8$). Figure 8a shows the diamagnetic response in all these materials. $T_c$ shows an enhancement with increase in value of *y*. Maximum $T_c$ of 4.4 K is observed for $y = 0.8$. This enhancement in $T_c$ can be attributed to unit cell contraction and local structure distortion as studied by *Chen et al*.[29]

Resistivity measurements, figure 8b, reconfirm the superconducting behavior in these materials. We should point out that in all the resistivity plots, a 'semiconductor'–like increase in resistivity is observed just before the beginning of the superconducting drop (Inset of Fig. 8b). This is commonly observed in high-$T_c$ cuprates and in F-doped LnOBiS$_2$ materials. In the upper panel of figure 9, we show the value of $T_c$ in these materials as measured magnetically and by resistivity. As already discussed above the pure Sm sample ($y = 1$) is non-superconducting till 2K (the limit of our measurement) which means that there may be a critical Sm concentration ($0.8 < y < 1$) which gives a maximum $T_c$ and further addition of Sm would suppress the superconducting state. Similar type of study has been very recently reported by *Kajitani et al*,[21] where Sm has been substituted for Nd in NdO$_{0.5}$F$_{0.5}$OBiS$_2$ and a maximum $T_c$ is achieved with Sm = 0.6, beyond which the $T_c$ suddenly diminishes. *Kajitani et al,*[21] also mentioned that there is a solubility limit for Sm in Nd$_{1-x}$Sm$_x$O$_{0.5}$F$_{0.5}$BiS$_2$ ($x$ = 0.8) but in our study La and Sm were completely soluble.

**Structure-property relationship**

Band structure calculations have revealed that the DOS near the Fermi level for these Bi-S superconductors is mainly due to Bi–6$p$ states with only a small contribution from S–3$p$ *states.*[33-34] The in-plane Bi–6$p$ orbitals are mainly involved in the overlap with S–3$p$ orbitals and thus, the in-plane (*ab*–plane) Bi–S bond lengths and the S–Bi–S bond angle seems to be important in inducing and enhancing superconductivity. Therefore the local structure in the Bi–S layers is an important parameter for superconductivity in LnOBiS$_2$ *vis-à-vis* FeAs layers in LnOFeAs superconductors. In order to get an insight, the effect of Sm substitution on local structure of Bi–S layers was investigated. The in-plane Bi–S$_{eq}$ distances and the in-plane S–Bi–S bond angle were calculated from structural data obtained after Rietveld analysis. The in–plane Bi–S bond lengths largely depend on the *a*-lattice parameter. In case of Sm doping at La sites the *a*–lattice parameter decrease gradually which leads to a decrease in the Bi–S distances as evident in lower panel of figure 9. Due to the decrease in the Bi–S distance a better overlap is expected which seems to enhance $T_c$. This trend is followed by all the Sm doped compositions. The in plane S–Bi–S angle is also expected to play a significant role in superconductivity. The flatter the bond (~180°) the better would be the overlap between the Bi–6$p$ and S–3$p$ orbitals. But as observed by *Chen, et al,*[29] a little distortion in the Bi–S layer could enhance $T_c$. A similar trend seems to follow in La$_{1-y}$Sm$_y$O$_{0.5}$F$_{0.5}$BiS$_2$. In case of $y$ = 1.0 *i.e.* SmO$_{0.5}$F$_{0.5}$BiS$_2$ composition the Bi–S bond length is slightly larger than other Sm–doped compositions which seems to overshadow the favorable effect of distortion in S–Bi–S bond angle. This may possibly be the reason for absence of superconductivity in $SmO_{0.5}F_{0.5}BiS_2$ but it needs to be further investigated.

## Conclusions

We have successfully synthesized $SmO_{1-x}F_xBiS_2$ which is a new member of the F–doped $LnOBiS_2$ family. This material is non superconducting and paramagnetic down to 2 K. We have investigated the structure and properties of the Sm substituted materials $La_{1-y}Sm_yO_{0.5}F_{0.5}BiS_2$ ($y$ = 0.0 – 1.0) with $Sm^{3+}$ ions smaller in size as compared with $La^{3+}$-ions and thus simulating chemical pressure as one effect of the substitution. Structural distortion in the Bi–S plane was analyzed. We find that the Bi–S bond lengths decrease with the increase in the doping level of the smaller Sm ion at the La site. We have observed superconductivity in $La_{1-y}Sm_yO_{0.5}F_{0.5}BiS_2$ ($y$ = 0.0 – 0.8) and we find that $T_c$ is gradually enhanced upon doping Sm at the La site as $y \rightarrow$ 0.8. Absence of superconductivity in $SmO_{0.5}F_{0.5}BiS_2$ is anomalous and deserves to be investigated. We also think that a tuning between these parameters via suitable doping or applied pressure could lead to superconductivity in $SmO_{1-x}F_xBiS_2$.

## ASSOCIATED CONTENT

**Supporting Information:**

Crystallographic Information File (CIF), powder X-ray diffraction patterns, results of Rietveld refinement and EDAX spectra are given in the supporting information. This material is available free of charge via the Internet at http://pubs.acs.org.

## AUTHOR INFORMATION

**\* Corresponding author**

Ashok Kumar Ganguli
Professor, Department of Chemistry
Indian Institute of Technology
New Delhi, 110016, India
Office Ph: 011 2659 1511
Email: ashok@chemistry.iitd.ac.in

**Present address:**
Ashok Kumar Ganguli
Director, Institute for Nano Science and Technology
Habitat Center, Phase X, sector 64
Mohali, Punjab – 160062

**Notes**

[†]Visiting scientist at Solid State and Nanomaterials Research Laboratory, Department of Chemistry, IIT Delhi, India.

The authors declare no competing financial interest.

## Acknowledgments

AKG acknowledges DST for financial support. GST and ZH acknowledge CSIR and UGC respectively for a fellowship. GKS and SA acknowledge the Department of Science and Technology (SERB & TSDP), CEFIPRA – New Delhi, UGC (BSR-Meritorious fellowship, SAP, MRP), for their financial support. Authors thank to Prof. A. Thamizhavel, TIFR (Mumbai) for provided the PPMS-VSM system for the magnetic measurements. Authors at IIT Delhi thank DST for providing the SQUID facility.

# FIGURES AND TABLES

**Figure 1.** Rietveld refinement studies of powder X-ray diffraction pattern for $SmO_{0.5}F_{0.5}BiS_2$. Pink and green vertical bars mark the allowed Bragg reflections for $SmO_{0.5}F_{0.5}BiS_2$ and $Bi_2S_3$ phases respectively. Inset shows the crystal structure of $Sm(O,F)BiS_2$.

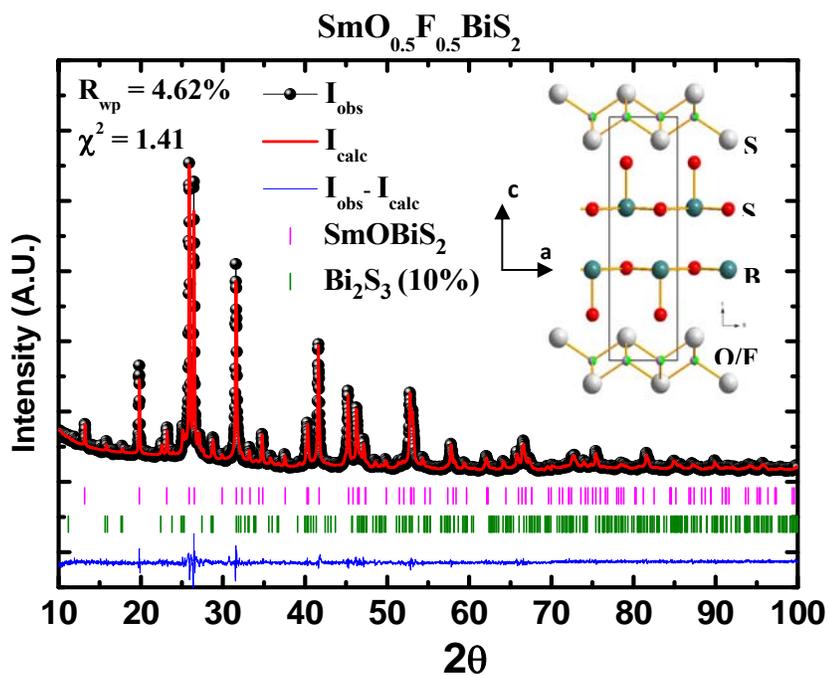

**Figure 2.** Variation of in-plane Bi–S bond length (Bi–$S_{eq}$) and S–Bi–S angle in known F–doped $LnOBiS_2$ and $Bi_4O_4S_3$.

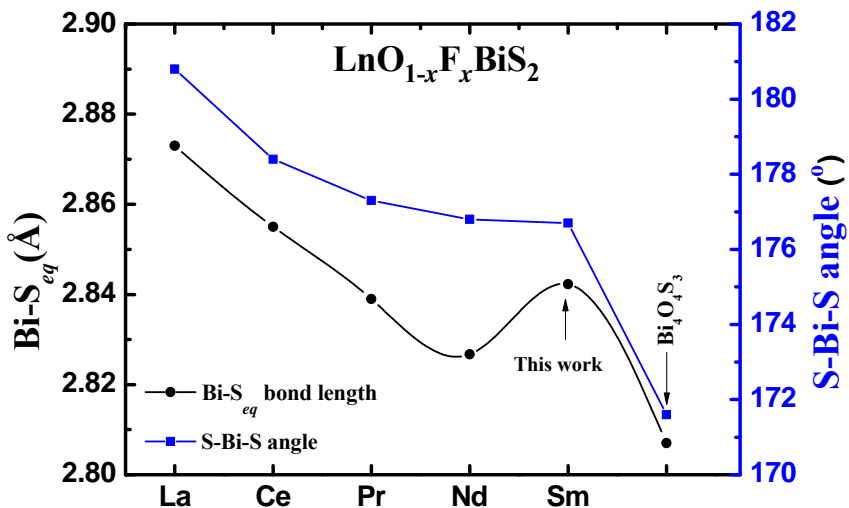

**Figure 3.** Variable temperature susceptibility data for SmO$_{1-x}$F$_x$BiS$_2$ (the sample comprises a collection of tiny single crystals) measured in a field of 30 Oe (main panel). Inset shows the isothermal magnetization data at different temperatures (5, 50, 150 and 300 K). The small hump at ~50 K is due to the presence of traces of frozen oxygen impurity

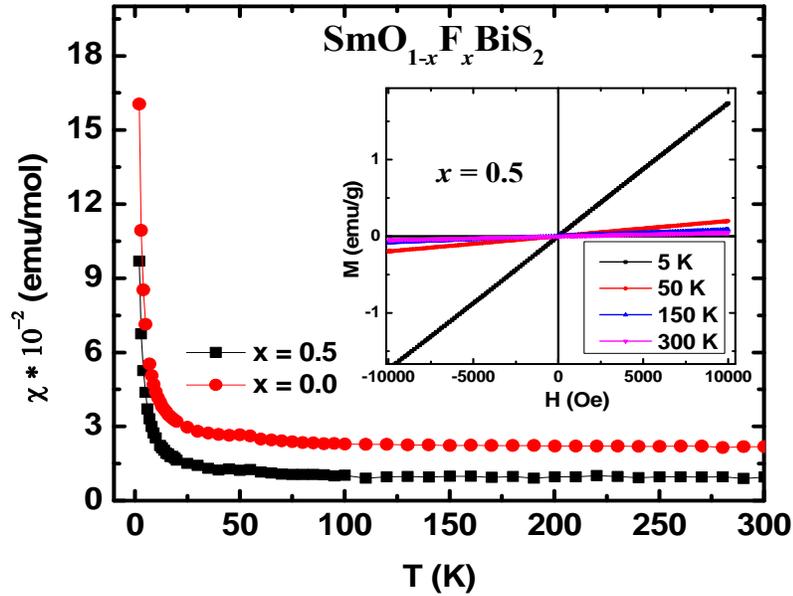

**Figure 4.** Variable temperature susceptibility and inverse susceptibility plot for SmO$_{0.5}$F$_{0.5}$BiS$_2$ (polycrystalline) in the low temperature region. Red line shows the Curie-Weiss fit.

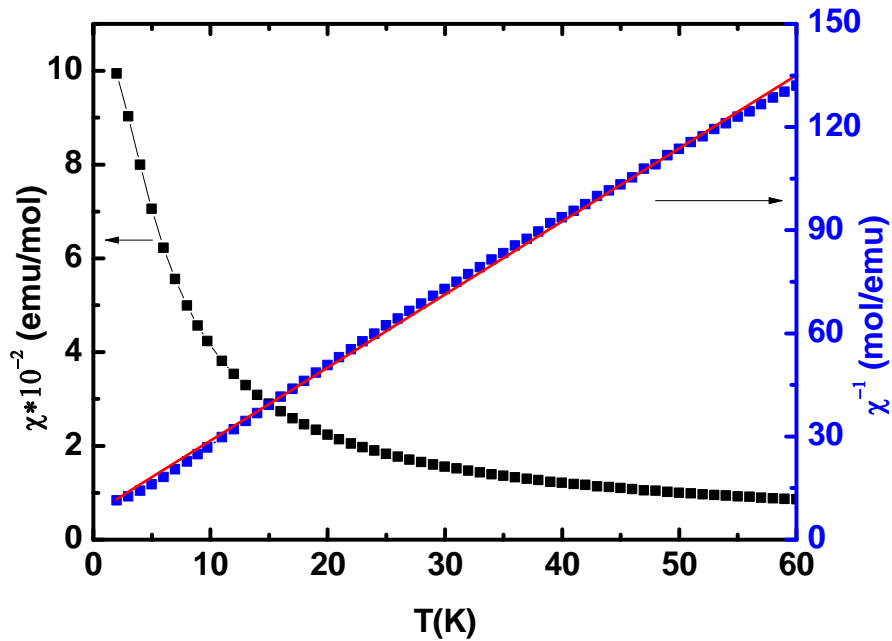

**Figure 5.** Resistivity versus temperature plots at different magnetic fields for $SmO_{0.5}F_{0.5}BiS_2$ (main panel) and for undoped $SmOBiS_2$ (inset).

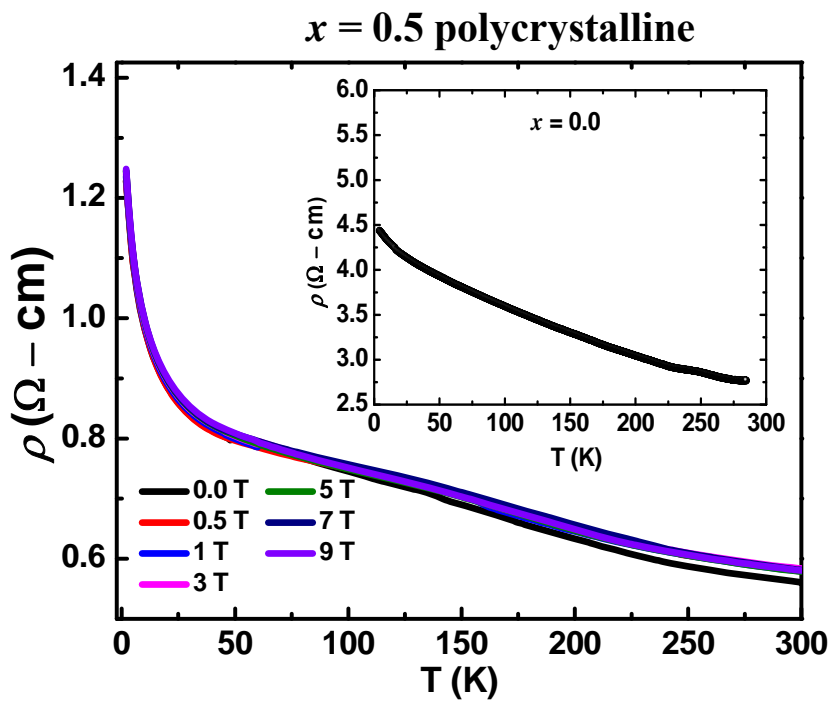

**Figure 6.** Powder X-ray diffraction patterns for $La_{1-y}Sm_yO_{0.5}F_{0.5}BiS_2$.

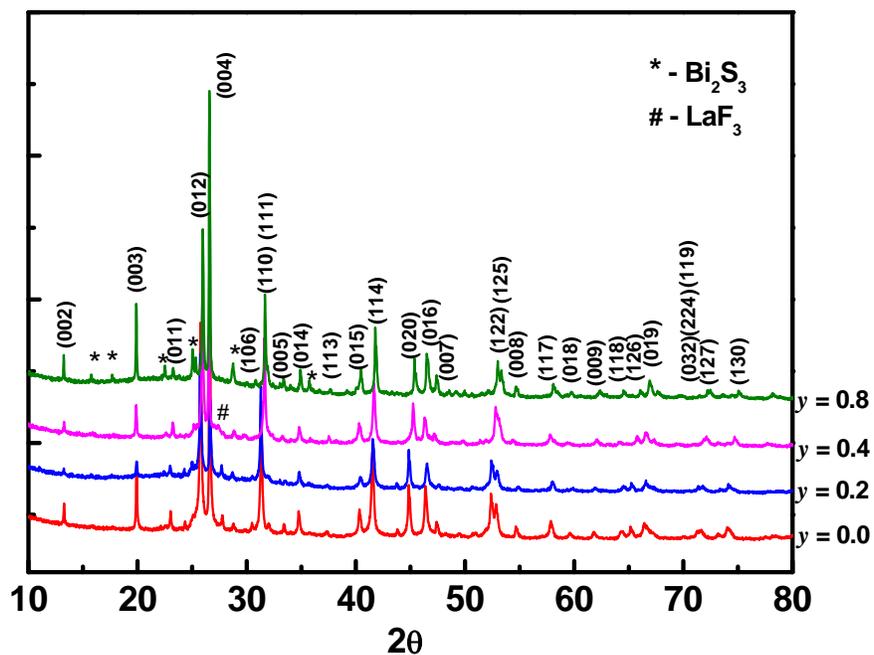

**Figure 7.** Variation of lattice parameter *a* and *c* with Sm content (*y*) for $La_{1-y}Sm_yO_{0.5}F_{0.5}BiS_2$.

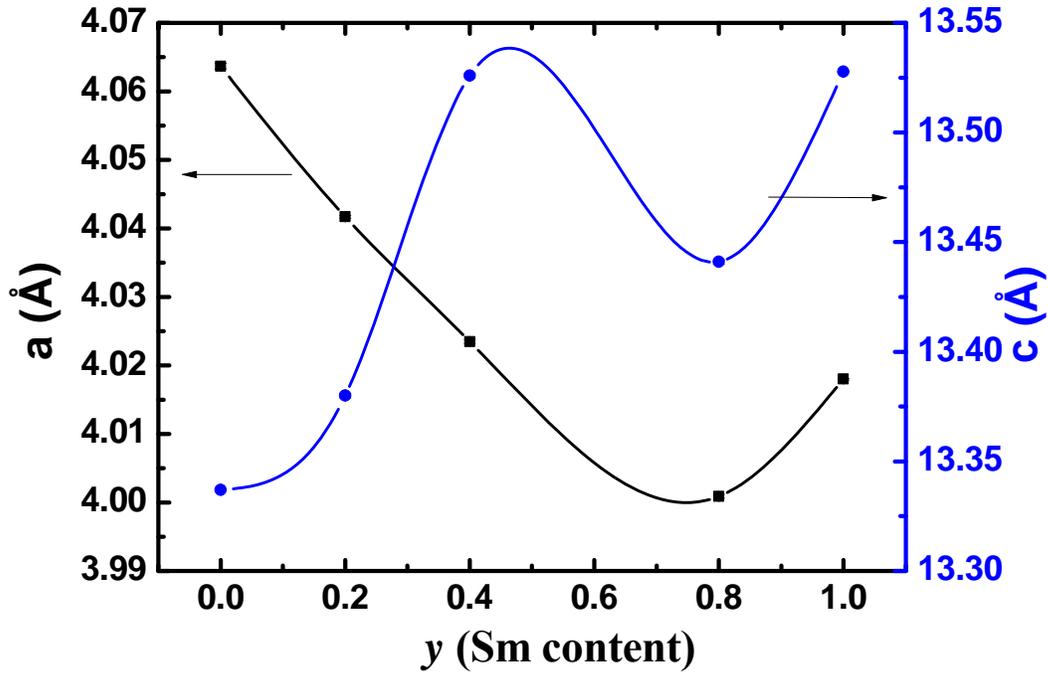

**Figure 8.** Variable temperature (a) magnetization and (b) resistivity curves for $La_{1-y}Sm_yO_{0.5}F_{0.5}BiS_2$. The magnetization was measured in ZFC condition in a field of 10 Oe. Inset of (b) show resistivity curves upto 300K.

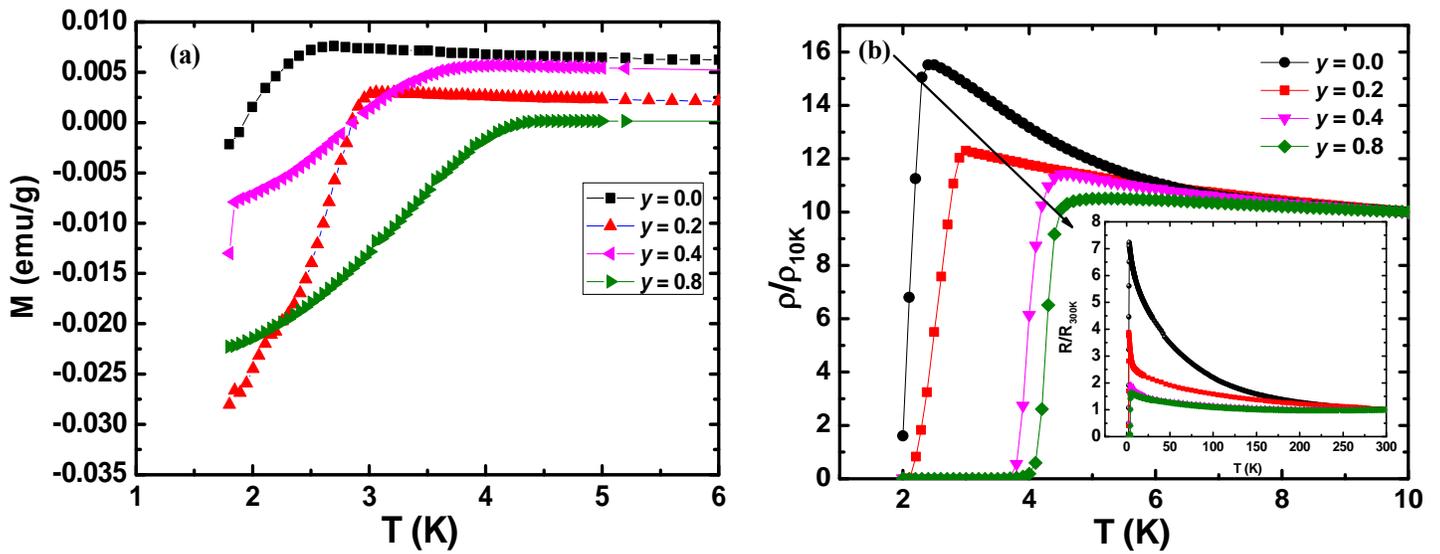

**Figure 9.** Plot of variation of $T_c$ obtained from magnetization and resistivity measurements (upper panel) and variation of Bi–S$_{eq}$ bond length and S–Bi–S angle with $y$ (lower panel) for La$_{1-y}$Sm$_y$O$_{0.5}$F$_{0.5}$BiS$_2$. The sample SmO$_{0.5}$F$_{0.5}$BiS$_2$ ($y$ = 1) does not exhibit superconductivity down to 2 K, the lowest temperature of our measurements.

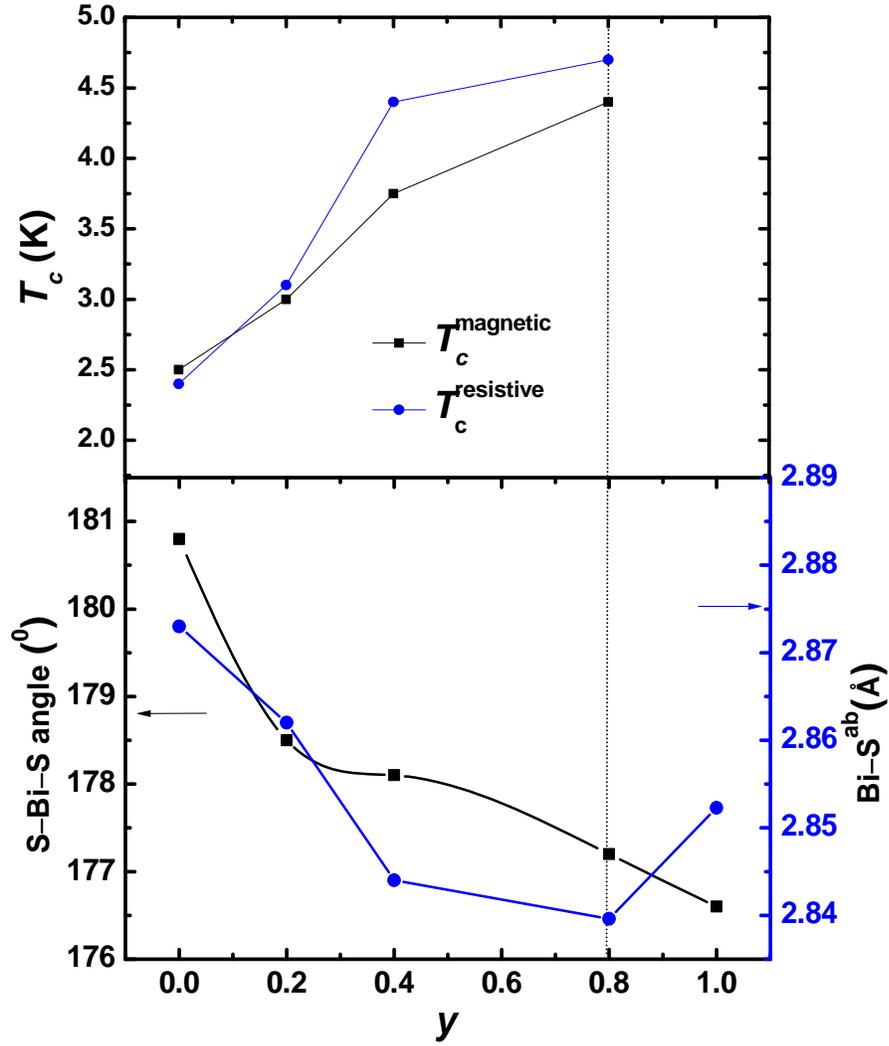

**Table 1.** Some crystal data and structural refinement parameters for $SmO_{0.5}F_{0.5}BiS_2$.

| Empirical formula | $SmO_{0.5}F_{0.5}BiS_2$ |
|---|---|
| Formula weight (g) | 440.95 |
| Space group, $Z$ | *P*4/*nmm* |
| Unit cell dimensions (Å) | $a$ = 4.018(1), $c$ = 13.534(3) |
| Volume (Å$^3$) | 218.50(1) |
| Density (g cm$^{-3}$) | 6.70 |
| Absorption coefficient (mm$^{-1}$) | 54.28 |
| Theta range (degree) | 1.5 – 28.8 |
| Index ranges | $-5 \leq h \leq 3$; $-5 \leq k \leq 5$; $-18 \leq l \leq 18$ |
| Reflections collected | 1096 |
| Indep. reflections | 213 |
| Data / parameters | 213/15 |
| Goodness-of-fit on $F^2$ | 1.17 |
| Final $R$ indices [$I$>2σ($I$)] | $R_1$ = 0.0530; $wR_2$ = 0.1143 |
| $R$ indices (all data) | $R_1$ = 0.0662; $wR_2$ = 0.1203 |
| Largest diff. peak and hole (eÅ$^{-3}$) | 3.94 and −2.93 |

**Table 2.** Atomic coordinates, Wyckoff positions, and isotropic equivalent displacement parameters for $SmO_{0.5}F_{0.5}BiS_2$ single crystal.

| Atom | Wyckoff | Occupancy | x | y | z | $U_{eq}$ (Å$^2$)$^a$ |
|---|---|---|---|---|---|---|
| Sm | 2c | 1 | 0.25 | 0.25 | 0.0949(2) | 0.0196(3) |
| Bi | 2c | 1 | 0.25 | 0.25 | 0.6262(1) | 0.0205(3) |
| S1 | 2c | 1 | 0.25 | 0.25 | 0.3799(4) | 0.022(4) |
| S2 | 2c | 1 | 0.25 | 0.25 | 0.8121(4) | 0.015(3) |
| O | 2a | 0.5 | 0.75 | 0.25 | 0 | 0.014(3) |
| F | 2a | 0.5 | 0.75 | 0.25 | 0 | 0.014(3) |
| Bi-S bond length Å ||||||||
| Equatorial (In-plane)[x4] = 2.8423(2) |||| Axial [x1] = 2.5181(2) |||
| In-plane S-Bi-S bond angle (x2) 176.69(4)° |||||||

$^a U_{eq}$ is defined as one-third of the trace of the orthogonalized $U^{ij}$ tensor.

Numbers in the parentheses indicate standard deviation.

# Supporting Information

## Synthesis and properties of $SmO_{0.5}F_{0.5}BiS_2$ and enhancement in $T_c$ in $La_{1-y}Sm_yO_{0.5}F_{0.5}BiS_2$

Gohil Singh Thakur[§], Ganesan Kalai selvan[¶], Zeba haque[§], Laxmi Chand Gupta[§†], Saroj Lochan Samal[‡], Sonachalam Arumugam[¶] and Ashok Kumar Ganguli[§]*.

[§]Solid State and Nano Research Laboratory, Department of Chemistry, Indian Institute of Technology New Delhi 110016, India.

[¶]Centre for High Pressure Research, School of Physics, Bharathidasan University, Tiruchirapalli 620024, India

[‡]Department of Chemistry, National Institute of Technology, Rourkela, Odisha, 769008, India

**CONTENTS:**

Powder X-ray diffraction data for $SmO_{1-x}F_xBiS_2$.

Results of Rietveld analysis on powder diffraction data for $SmO_{0.5}F_{0.5}BiS_2$.

SEM and EDAX analysis for $SmO_{1-x}F_xBiS_2$ crystals.

Results of Rietveld refinement to the powder X-ray diffraction data for $La_{1-y}Sm_yO_{0.5}F_{0.5}BiS_2$ ($y$ = 0.0, 0.2, 0.4 and 0.8).

Magnified X-ray diffraction (XRD) patterns of $La_{1-y}Sm_yO_{0.5}F_{0.5}BiS_2$ in the region around (004) and (200) peaks.

**Table S1. Anisotropic displacement parameters of $SmO_{0.5}F_{0.5}BiS_2$ obtained from single crystal refinement.**

| Atom | U11 | U22 | U33 | U23 | U13 | U12 |
|---|---|---|---|---|---|---|
| Sm | 0.0170(8) | 0.0170(8) | 0.025(1) | 0.0 | 0.0 | 0.0 |
| Bi | 0.0203(6) | 0.0203(6) | 0.0210(9) | 0.0 | 0.0 | 0.0 |
| S1 | 0.021(4) | 0.021(4) | 0.035(5) | 0.0 | 0.0 | 0.0 |
| S2 | 0.012(3) | 0.012(3) | 0.021(4) | 0.0 | 0.0 | 0.0 |
| F/O | 0.005(1) | 0.005(1) | 0.034(9) | 0.0 | 0.0 | 0.0 |

**Table S2. Atomic coordinates and Wyckoff positions for $SmO_{0.5}F_{0.5}BiS_2$ obtained after Rietveld analysis of powder diffraction data.**

The occupancies for all the atoms were refined initially but were found to be above 97% and hence were then fixed to unity.

| Space group: *P4/nmm* <br> a = 4.0180(1) Å and c = 13.5291(4) Å | | | | | | |
|---|---|---|---|---|---|---|
| Atom | Wyckoff | Occ. | x | y | z | $B_{iso}$ |
| Sm | 2c | 1 | 0.25 | 0.25 | 0.0949(1) | 0.51(3) |
| Bi | 2c | 1 | 0.25 | 0.25 | 0.6226(7) | 0.43(1) |
| S1 | 2c | 1 | 0.25 | 0.25 | 0.3856(3) | 0.86(2) |
| S2 | 2c | 1 | 0.25 | 0.25 | 0.8081(3) | 1.5(1) |
| O/F | 2a | 0.5/0.5 | 0.75 | 0.25 | 0 | 1.9(3) |
| In-plane S-Bi-S angle = 175.5° <br> In-plane Bi-S bond length ( x 4) = 2.843(4) Å <br> $R_{wp}$ (%) = 4.62; $\chi^2$ = 1.41 | | | | | | |

**Figure S1.** Powder X-ray diffraction data for $SmO_{1-x}F_xBiS_2$ polycrystalline and crushed single crystal samples.

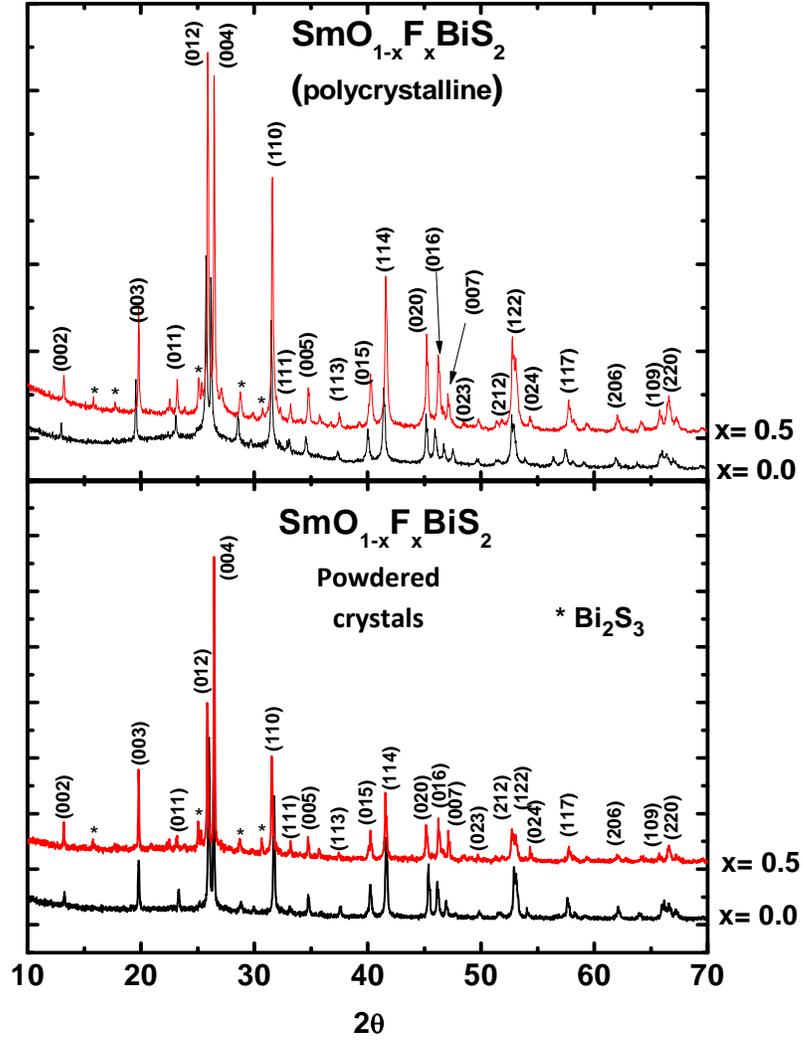

**Figure S2.** Electron micrograph of a typical SmO$_{0.5}$F$_{0.5}$BiS$_2$ crystal with dimension (220 μm x 120 μm x 20 μm).

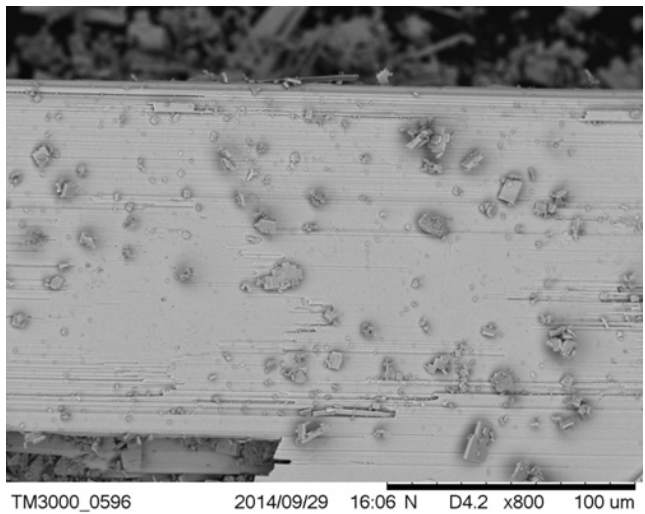

Many crystals of different sizes with one of the longest dimension ranging from 20 – 300 $\mu$m were obtained. The crystals were not big enough to be fit for Resistivity and magnetization measurements and hence for resistivity a sintered pellet of polycrystalline sample was used. For magnetization measurements a mixture of many tiny crystals of different sizes was used.

**Figure S3.** Results of SEM-EDAX analysis on a selected SmO$_{0.5}$F$_{0.5}$BiS$_2$ single crystal.

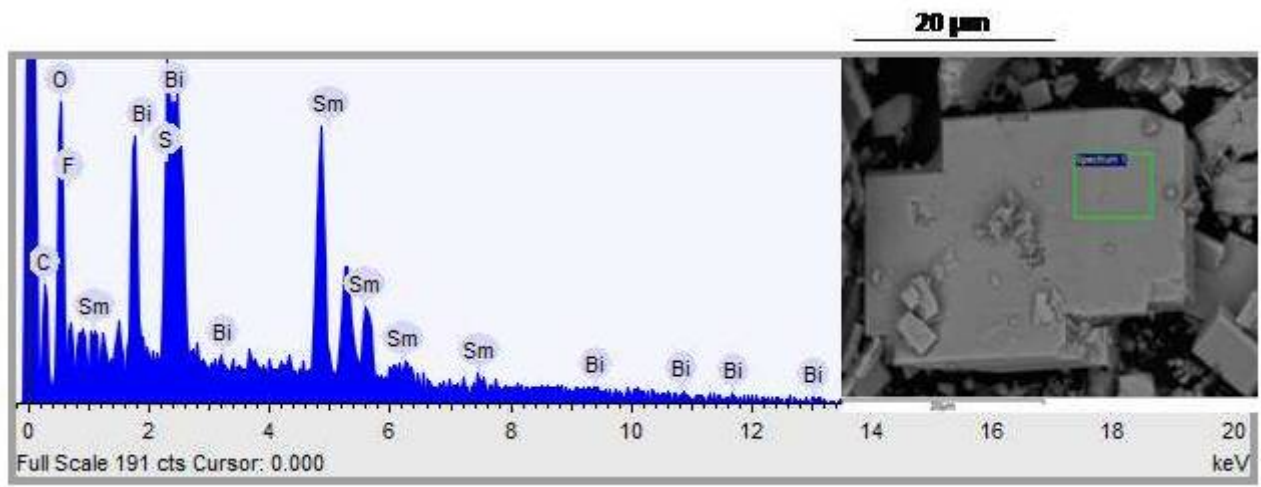

| Element | F | S | Sm | Bi |
|---------|-----|-------|-------|-------|
| Atom %  | 8.1 | 42.95 | 20.81 | 19.96 |

Presence of all the constituent elements is evident from the EDAX graph. The table denotes the percentage composition of each element in the sample. EDAX data was collected on many such crystals and the final stoichiometry was deduced by averaging the data obtained from each point. The average approximate stoichiometry was calculated to be **SmBi$_{0.96}$S$_{2.1}$F$_{0.39}$**.

**Figure S4.** Results of Rietveld refinement of the powder X-ray diffraction data for La$_{1-y}$Sm$_y$O$_{0.5}$F$_{0.5}$BiS$_2$ ($y$ = 0.0, 0.2, 0.4 and 0.8). A small amount of impurities of LaF$_3$, BiF$_3$, Bi$_2$O$_3$ and/or Bi$_2$S$_3$ were observed for all the compositions.

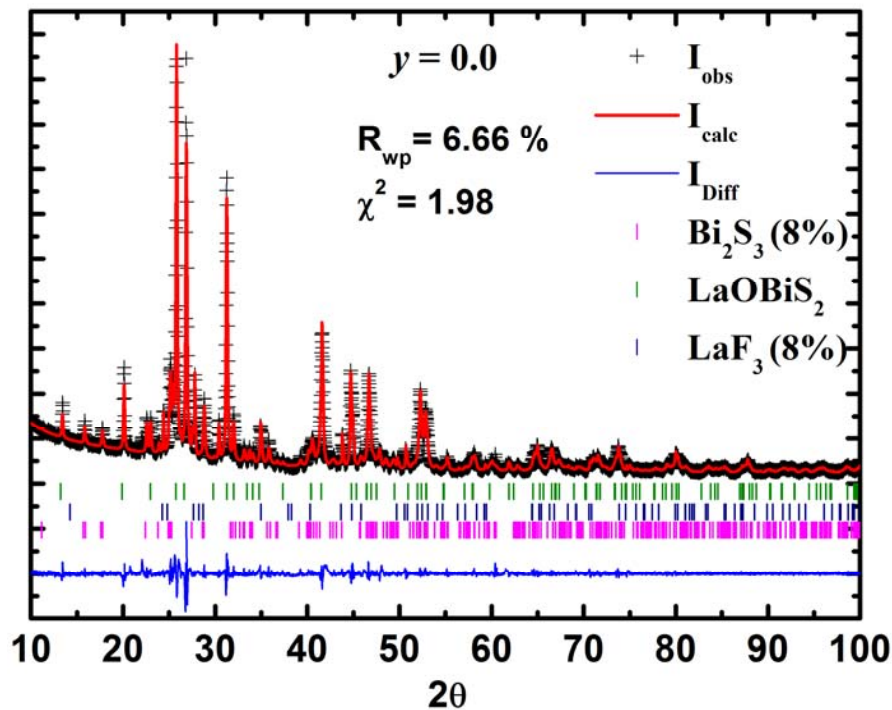

**y = 0.0**

$a$ = 4.0637(4) Å

$c$ = 13.3367(4) Å

| Site | x | y | z | Atom | Occ | B$_{eq}$ |
|---|---|---|---|---|---|---|
| 2c | 0.25 | 0.25 | 0.1005(2) | La | 1 | 0.55(2) |
| 2c | 0.25 | 0.25 | 0.6212(1) | Bi | 1 | 0.35(2) |
| 2c | 0.25 | 0.25 | 0.3937(3) | S | 1 | 0.90(4) |
| 2c | 0.25 | 0.25 | 0.8046(2) | S | 1 | 1.20(7) |
| 2a | 0.75 | 0.25 | 0.0 | O/F | 0.5/0.5 | 1.9(3) |

**In-plane S-Bi-S angle = 180.8 °**

**In-plane Bi-S bond length (x 4) = 2.872 Å**

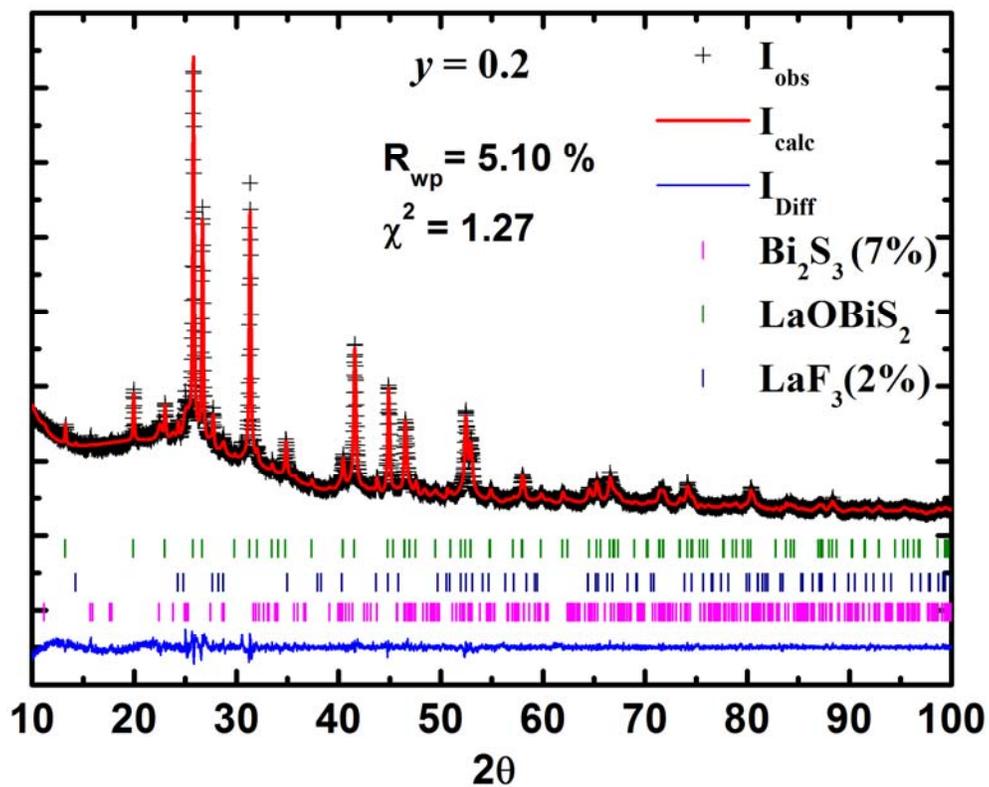

**y = 0.2**

a = 4.0417(2) Å

c = 13.3839(3) Å

| Site | x | y | z | Atom | Occ | $B_{iso}$ |
|---|---|---|---|---|---|---|
| 2c | 0.25 | 0.25 | 0.0959(8) | Sm | 0.222(3) | 0.61(1) |
| 2c | 0.25 | 0.25 | 0.0959(8) | La | 0.778(3) | 0.61(2) |
| 2c | 0.25 | 0.25 | 0.6155(6) | Bi | 1 | 0.35(2) |
| 2c | 0.25 | 0.25 | 0.3861(3) | S1 | 1 | 0.52(6) |
| 2c | 0.25 | 0.25 | 0.8125(3) | S2 | 1 | 1.52(5) |
| 2a | 0.75 | 0.25 | 0.0 | O/F | 0.5/0.5 | 1.4(3) |

**In-plane S-Bi-S angle = 178.5 °**

**In-plane Bi-S bond length (x 4) = 2.860 Å**

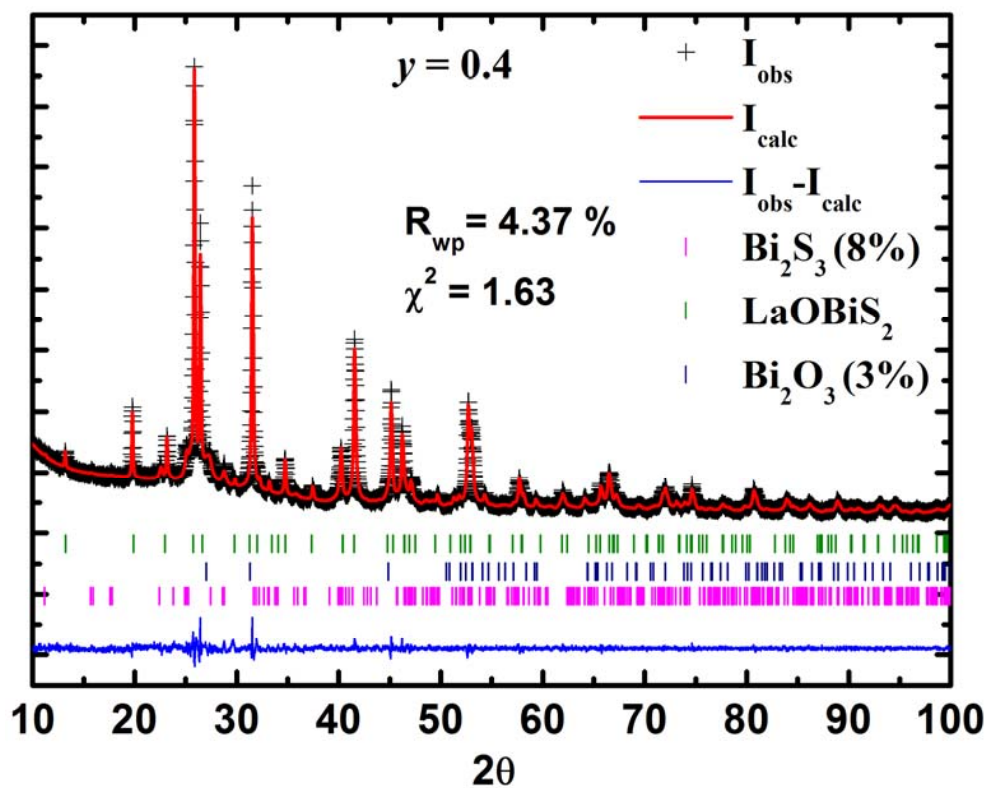

**y = 0.4**

a = 4.0235(2) Å

c = 13.5261(1) Å

| Site | x | y | z | Atom | Occ | $B_{iso}$ |
|------|------|------|-----------|------|----------|---------|
| 2c | 0.25 | 0.25 | 0.0938(2) | Sm | 0.385(1) | 0.43(7) |
| 2c | 0.25 | 0.25 | 0.0938(2) | La | 0.615(1) | 0.43(7) |
| 2c | 0.25 | 0.25 | 0.6245(1) | Bi | 1 | 0.38(2) |
| 2c | 0.25 | 0.25 | 0.3790(1) | S1 | 1 | 0.67(6) |
| 2c | 0.25 | 0.25 | 0.8050(3) | S2 | 1 | 1.3(5) |
| 2a | 0.75 | 0.25 | 0.0 | O/F | 0.5/0.5 | 1.9(2) |

**In-plane S-Bi-S angle = 178.1 °**

**In-plane Bi-S bond length (x 4) = 2.844 Å**

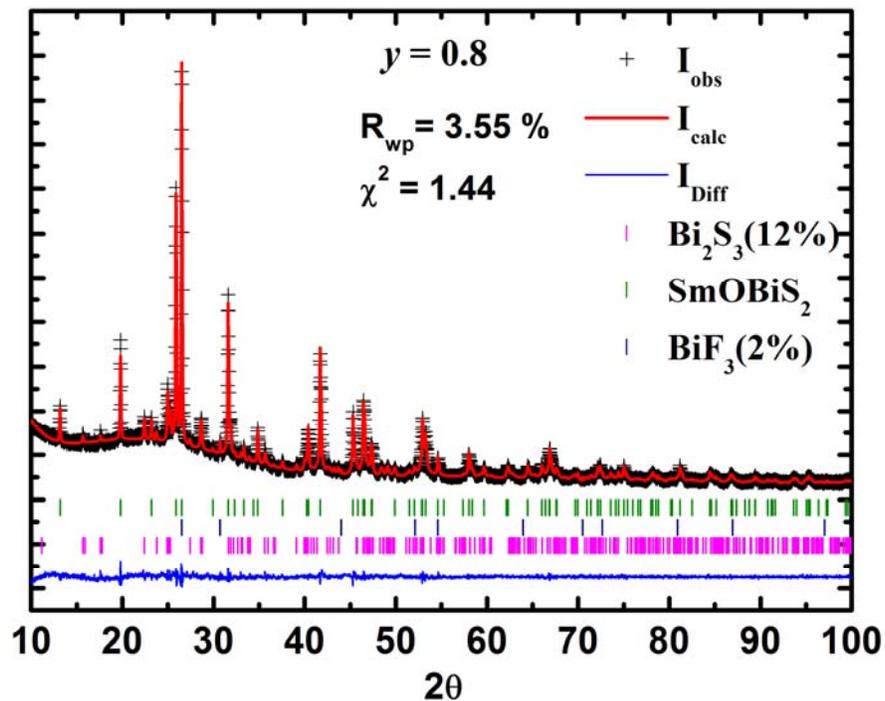

**y = 0.8**

$a$ = 4.0008(2) Å

$c$ = 13.4410(1) Å

| Site | $x$ | $y$ | $z$ | Atom | Occ | $B_{iso}$ |
|------|-----|-----|-----|------|-----|-----------|
| 2c | 0.25 | 0.25 | 0.1094(1) | Sm | 0.823(4) | 0.69(2) |
| 2c | 0.25 | 0.25 | 0.1094(1) | La | 0.177(4) | 0.69(2) |
| 2c | 0.25 | 0.25 | 0.6224(5) | Bi | 1 | 0.53(3) |
| 2c | 0.25 | 0.25 | 0.3786(1) | S1 | 1 | 0.83(5) |
| 2c | 0.25 | 0.25 | 0.8125(4) | S2 | 1 | 0.90(5) |
| 2a | 0.75 | 0.25 | 0.0 | O/F | 0.5/0.5 | 1.21(3) |

**In-plane S-Bi-S angle = 177.2 °**

**In-plane Bi-S bond length (x 4) = 2.836 Å**

**Figure S5.** Magnified X-ray diffraction (XRD) patterns of $La_{1-y}Sm_yO_{0.5}F_{0.5}BiS_2$ in the region around (004) and (200) peaks.

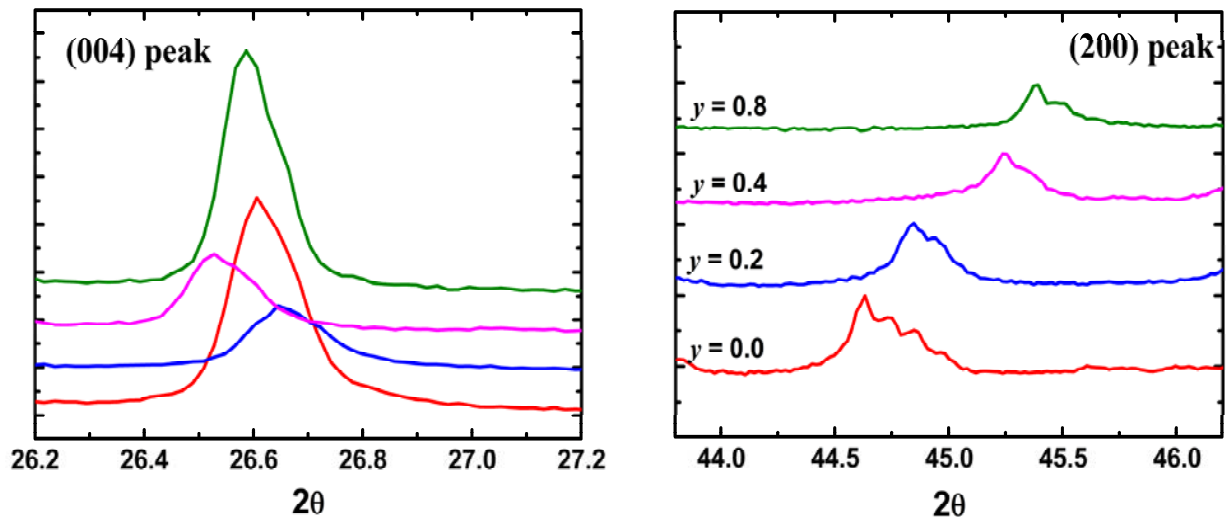

The regular shift of (200) peaks towards higher angle with increase in Sm doping indicates $a$-axis contraction. The (004) peaks do not show such regularity which indicates that the change in $c$-axis is not linear to the change in the Sm contraction.